\documentstyle[epsfig,graphicx]{aipproc}

\def\netmu{{\cal N}_{\mu}}
\def\DSTPM{{\rm D}^{\ast\pm}}
\def\DSTP{{\rm D}^{\ast+}}
\def\DSTM{{\rm D}^{\ast-}}
\def\DST{{\rm D}^{\ast}}
\def\D0{{\rm D}^0}
\newcommand{\sqee}{\sqrt{s_{\rm ee}}}
\newcommand{\epem}{$\mbox{e}^+\mbox{e}^-$}
\def\ptdst{p_{\rm T}^{\DST}}
\def\etadst{\eta^{\DST}}
\def\dspt{{\rm d}\sigma/{\rm d}p_{\rm T}^{\DST}}

\def\xgmin{x^{\rm min}_{\gamma}}
\def\xt{x_{\rm T}^{\DST}}
\def\DELTACAND{\Delta M \equiv M_{\DST}^{\rm cand}-M_{\D0}^{\rm cand}}
\def\DELTAM{\Delta M \equiv M_{\DST}-M_{\D0}}
\def\CF2{F_{\rm 2,c}^{\gamma}}

\def\ETJET{E_{\rm T}^{\rm jet}}
\def\ptrel{p_{\rm T,rel}^{\mu}}
\def\pth{p_{\rm T,rel}^{\rm h}}

\def\ccbar{\mbox{c}\overline{\mbox{c}}}

\def\c{\mbox{c}}
\def\b{\mbox{b}}
\def\uds{\mbox{uds}}
\def\BG{\rm BG}

\def\sigcc{\sigma({\rm e}^+{\rm e}^-\to{\rm e}^+{\rm e}^-{\rm c}\bar{{\rm c}})}
\def\sigbb{\sigma({\rm e}^+{\rm e}^-\to{\rm e}^+{\rm e}^-{\rm b}\bar{{\rm b}})}

\def\sigmeas{$\sigma_{\rm meas}^{\DST} = 30.7\pm 2.8({\rm stat})\pm 3.3({\rm
syst})$~pb}
\def\sigtot{1033\pm 102~({\rm stat})\pm  111~({\rm syst})\pm 
246~({\rm extr})~{\rm pb}}
\def\sigdir{362\pm 59~({\rm stat})\pm  42~({\rm syst})\pm
 68~({\rm extr})~{\rm pb}}
\def\sigres{671\pm 110~({\rm stat})\pm 77~({\rm syst})\pm
178~({\rm extr})~{\rm pb}}
\def\fb{(27.2 \pm 4.8~({\rm stat}))\%} 
 
\def\sb{\sigbb=14.2 \pm 2.5~({\rm stat})~_{+5.3}^{-4.8}~({\rm sys})~\mbox{pb}}

\begin{document}
\title{Charm and bottom production
in two-photon collisions with OPAL\protect\footnote{Talk given at the PHOTON 2000 Conference, Ambleside, 
        UK, August 2000.}}

\author{\'Akos Csilling\thanks{The author thanks the UK Particle Physics and Astronomy Research Council
for their support.\break  
Permanent address: KFKI Research Institute
for Particle and Nuclear Physics, Budapest,  P.O.Box 49, H-1525 Hungary;
supported by the Hungarian Foundation for Scientific Research, OTKA F-023259.}}
\address{University College London\\
Gower Street, London WC1E 6BT, UK
}

\null\vspace{-2mm}
\maketitle

\begin{abstract}
A preliminary update of the previous OPAL measurement of the inclusive
production of $\DSTPM$ mesons in anti-tagged  photon-photon collisions is
presented together with the first preliminary OPAL measurement of bottom
production in photon-photon collisions.
\end{abstract}

\section*{Introduction}

Heavy quark production in photon-photon collisions reveals the structure of a
fundamental gauge boson, the photon. It can be calculated in perturbative QCD
and is sensitive to the quark and gluon content of the photon. At LEP2 energies
the two main processes, direct and single-resolved, have approximately the same
contribution, while the double-resolved contribution is negligible. 


Charm production  in photon-photon collisions can be cleanly tagged e.g.~by the
full reconstruction of charged $\DST$ mesons, while bottom production can be
measured through a careful study of the various sources of final state leptons
in hadronic two-photon events.

\section*{Charm production}

We present a preliminary update on the measurement of $\DST$ production in
anti-tagged photon-photon collisions~\cite{bib-opaldstpaper}. 
The main change compared to the previous analysis is the inclusion of  data
collected by OPAL in 1999 at  \epem~centre-of-mass energies  $\sqee=192 -
202~{\rm GeV}$, bringing the total integrated luminosity to
$428~\mathrm{pb}^{-1}$ in the centre-of-mass energy range  $183<\sqee<202$ GeV
with a luminosity-weighted  average energy of
$\langle\sqee\rangle=193~\mathrm{GeV}$.

\begin{figure}[tbp]
   \begin{center}
    \includegraphics[width=0.51\textwidth]{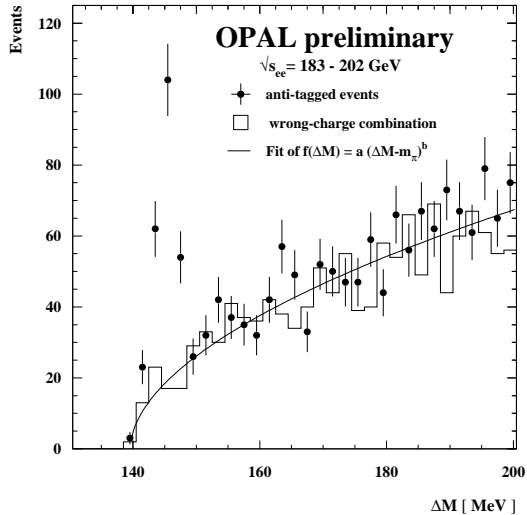}
   \end{center}
\caption{          
  Mass difference $\DELTACAND$ for the full sample. A clear peak is visible
  around $\DELTAM = 145.4~{\rm MeV}$. The open histogram represents the
  wrong-charge  sample which gives a good description  of the
  combinatorial background.  The result of a fit to the background 
is superimposed.
       }                    
\label{fig-signal}
\end{figure}

 $\DSTP$ mesons are reconstructed in their decay to ${\rm D}^0\pi^+$, with the
${\rm D}^0$ observed in the two decay modes ${\rm K}^-\pi^+$ (3-prong) and
${\rm K}^-\pi^+\pi^-\pi^+$ (5-prong). The charge-conjugate decay
chain is used for $\DSTM$ mesons. The distribution of the mass difference between the
$\DST$ and the $\D0$,  $\Delta M$, for the combined data set is shown in
Figure~\ref{fig-signal}.  Background is subtracted by fitting  the function $
f(\Delta M)=a \cdot (\Delta M-m_{\pi})^b$,  where $a$ and $b$ are  free 
parameters and $m_{\pi}$ is the pion mass, to the combined upper sideband of
the signal and  the wrong-charge distribution in the range $160.5~{\rm
MeV}<\Delta M<200.5~{\rm MeV}$. The result of the fit is in good agreement with
the wrong-charge combination in the signal region. In total we observe 
$164.3\pm 14.9~(\rm stat)$  $\DST$ mesons.

According to the PYTHIA Monte Carlo program the $\DST$ selection efficiencies
for the transverse momentum and rapidity range of 
$\ptdst>2~{\rm GeV}$ and $|\etadst|<1.5$ are around
$31-34\%$  for the 3-prong and $7-9\%$ for the 5-prong decay modes in 
single-resolved and direct events. The approximately 6\% lower efficiency
compared to the published analysis is due to additional electron rejection
cuts which significantly improve the purity of the signal.

\begin{figure}[tbp]
    \includegraphics[width=0.5\textwidth]{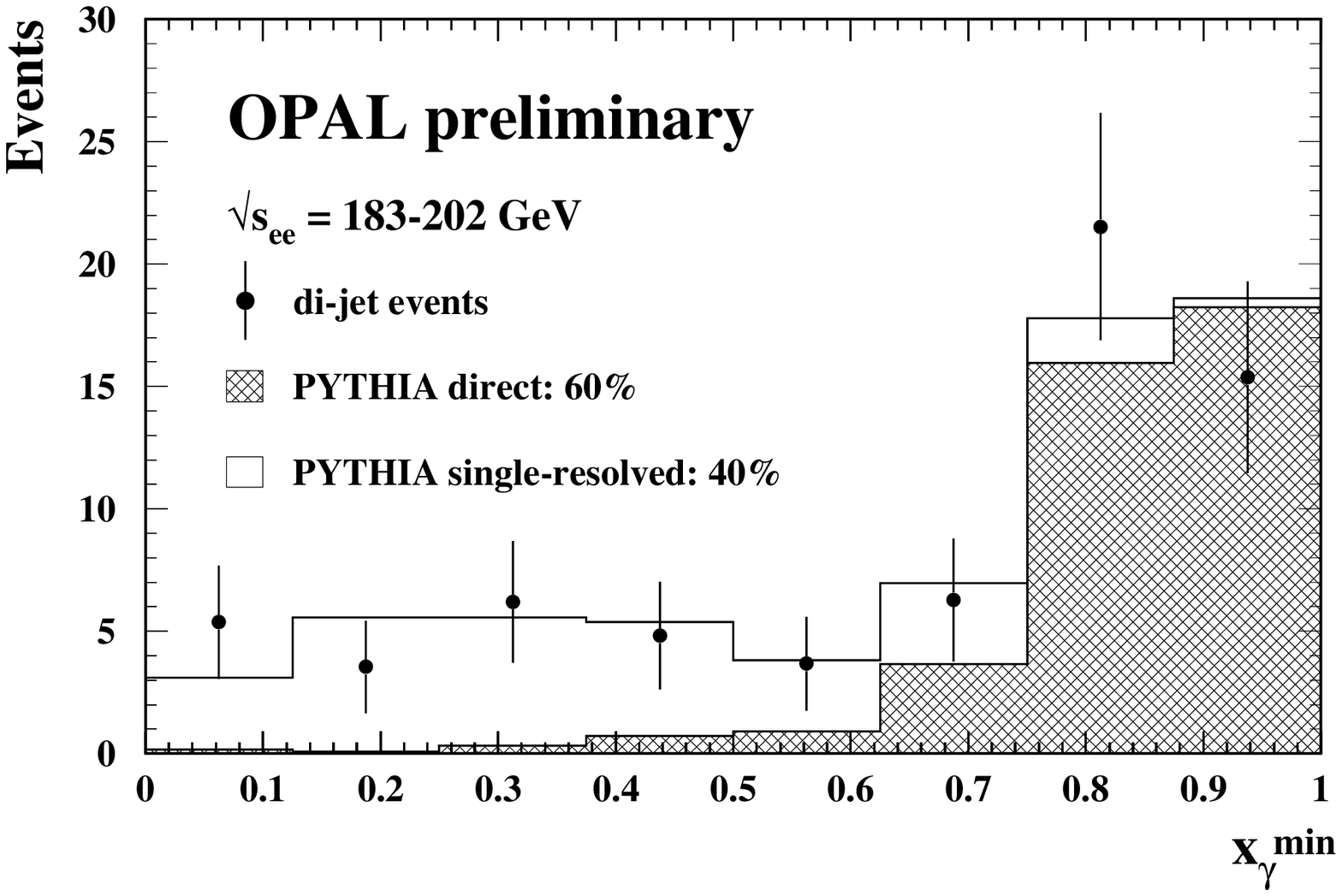}\hfill
    \includegraphics[width=0.5\textwidth]{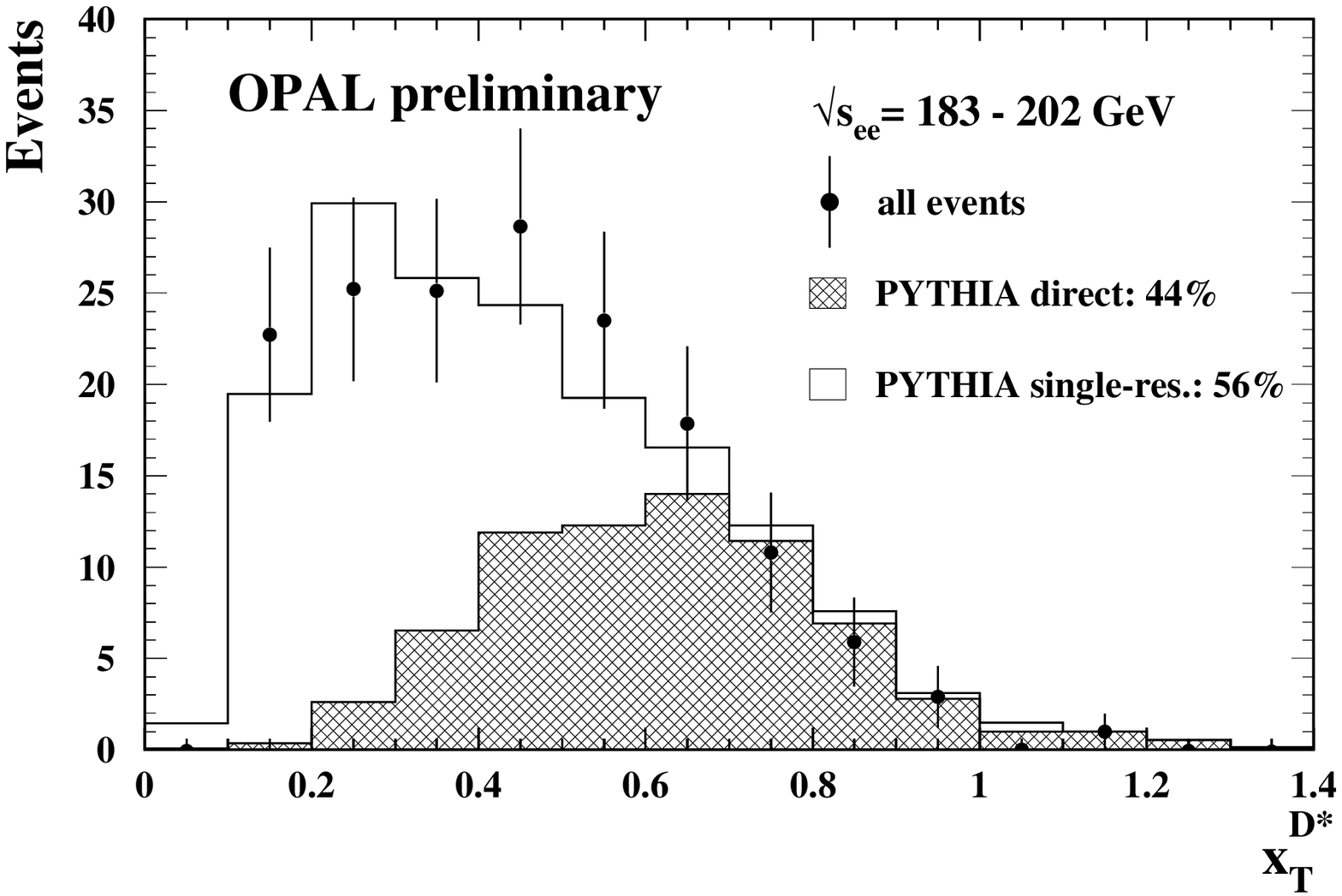}\vspace{1pt}
\caption{
         $x_{\gamma}^{\mathrm{min}}$  for dijet
         events (left) and  $\xt$ 
          for all events (right)
	 in the signal region.
         The open area shows the fitted contribution from the 
	 single-resolved process, and the hatched area
         the one from the direct process.}
\label{fig-xgamma}
\label{fig-ptdst}
\end{figure}

Two methods are  used to determine  the relative fraction of direct and 
single-resolved events in the data sample. One is based  on $\xgmin$, defined 
 in two-jet events using the cone algorithm as the minimum of 
$x_{\gamma}^{\pm}  = {\Sigma_{\rm jets}(E\pm p_z)}/{\Sigma_{\rm hadrons}(E\pm 
p_z)}$, while the other uses  the scaled $\DST$ transverse momentum 
$\xt={2\ptdst}/{W_{\rm vis}}$, available in all events. The results of both
methods are shown in Figure~\ref{fig-xgamma}. The ratio of direct to
single-resolved  contributions in the dijet events determined by a fit  to the
$\xgmin$ distribution yields  $(60\pm 8)\%$  direct and $(40\pm 8)\%$ 
single-resolved, while the fit to the $\xt$  distribution yields  $(44\pm
6)\%$  direct and $(56\pm 6)\%$   single-resolved contributions, where all
errors are statistical only. The requirement of two jets introduces a bias 
towards a larger direct component,
therefore the $\xt$ method, where the full sample is used, is preferred.

\begin{figure}[tbp]
\null\vspace{-5mm}
    \includegraphics[width=.495\textwidth]{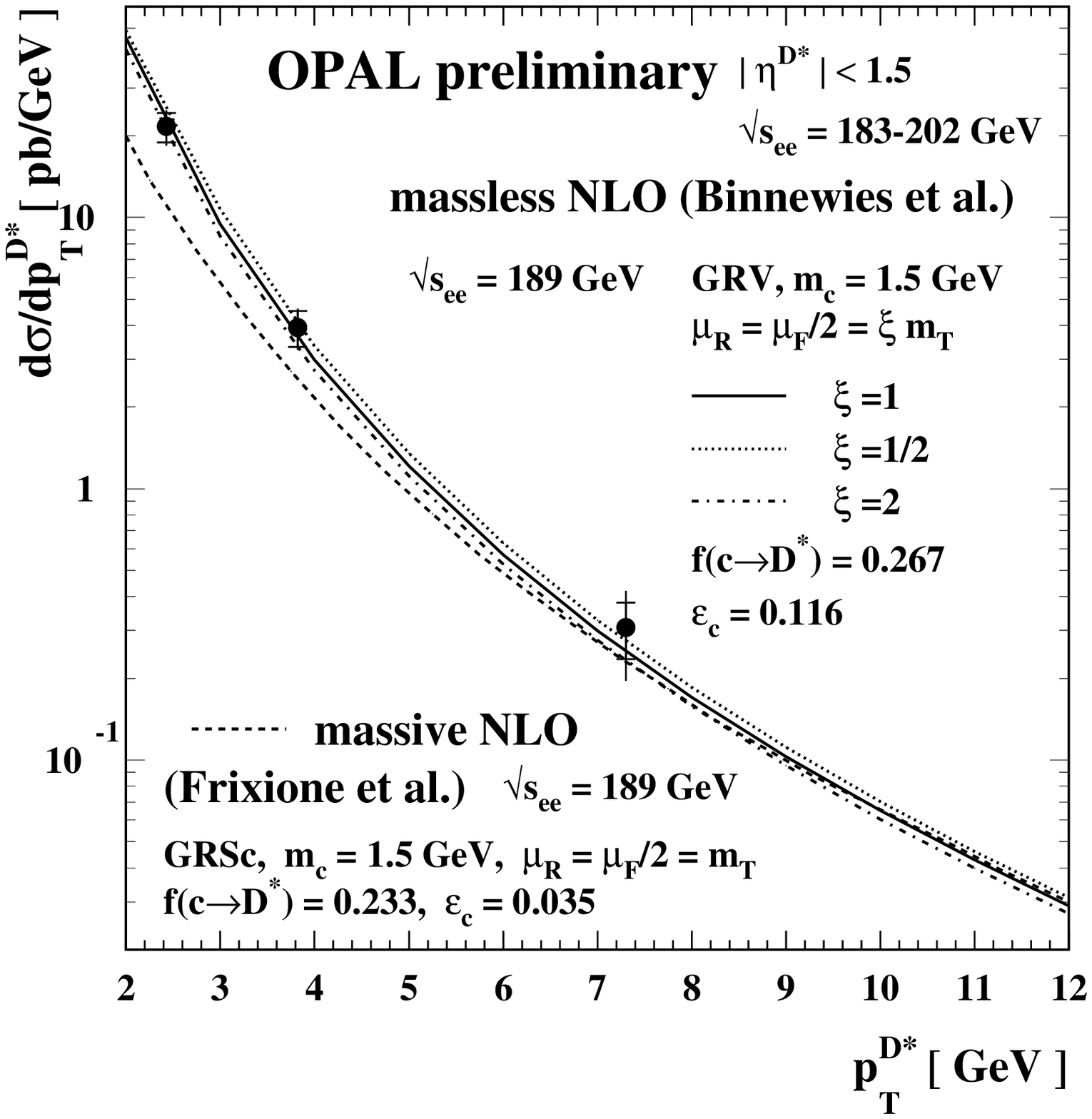}
    \hfill
    \includegraphics[width=.495\textwidth]{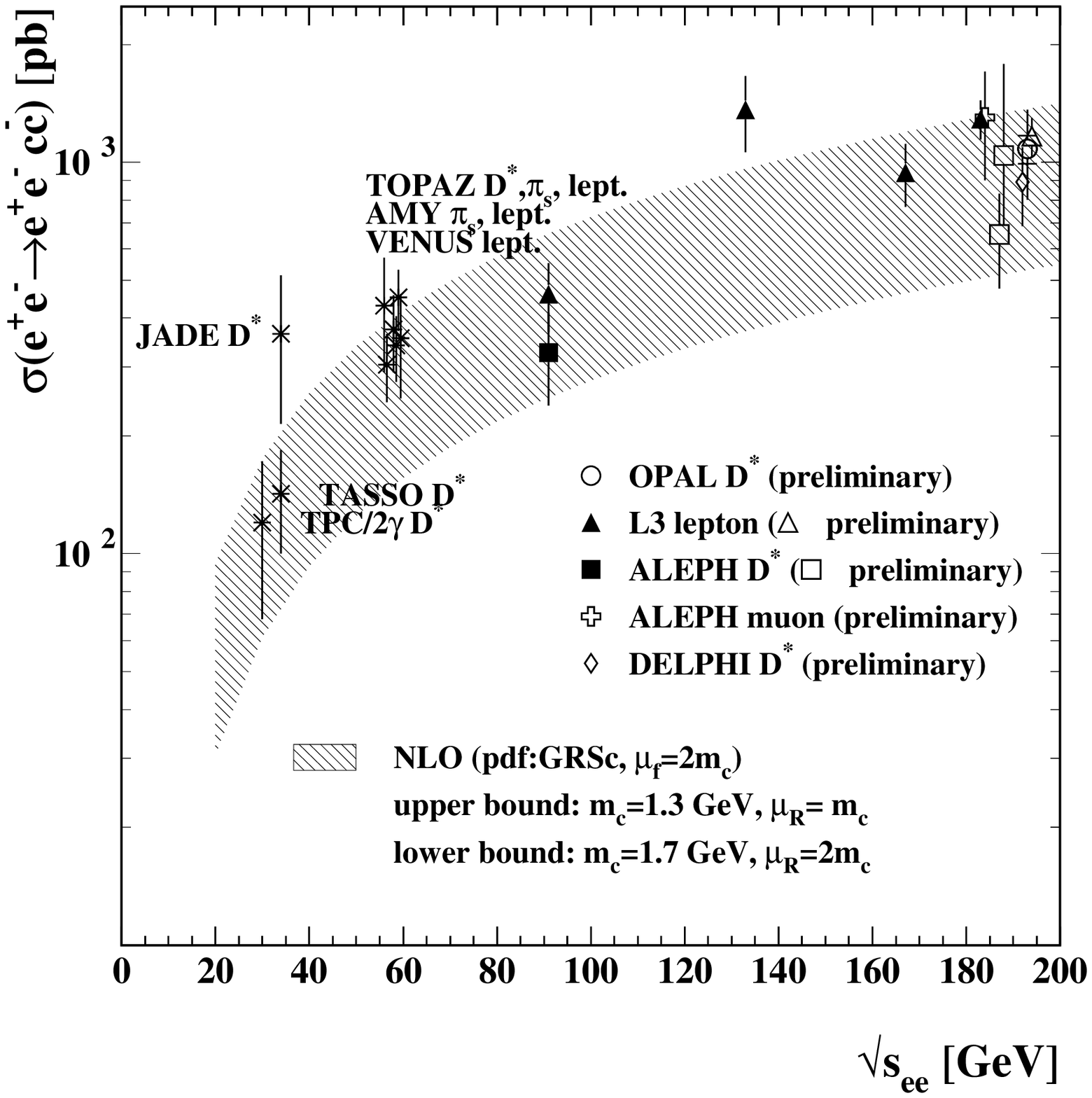}
\caption{ The differential $\DST$ cross-section, $\dspt$, for the process
         \epem $\to$ \epem $\DST X$
         in the range $|\etadst|<1.5$, 
	 compared to an NLO calculation  using the massless 
	 approach~\protect\cite{bib-Kniehl} 
         and another one using the
         massive approach~\protect\cite{bib-Frixione} (left) 
	 and  the total cross-section for the process
         \epem $\to$ \epem $\ccbar$ also compared to
          an NLO calculation~\protect\cite{bib-DKZZ} and other
	  measurements~\protect\cite{bib-data}~(right).} 
\label{fig-sigpt}
\label{fig-sigtot}
\end{figure}

In Figure~\ref{fig-sigpt} the combined differential cross-section $\dspt$ is
compared to a next-to-leading order (NLO) calculation by Frixione et 
al.~\cite{bib-Frixione}, where the matrix elements
for massive charm quarks are used,  and to an NLO calculation by Binnewies
et~al.~\cite{bib-Kniehl}, where charm is treated as
a massless, active flavour in the photon parton distribution function.
%
%
Despite the low
transverse momenta studied,  the massless calculation describes the
measurement, while the massive calculation underestimates it in the region of
small $\ptdst$. This is contrary to the expectation that the massive approach
should be more  appropriate at lower transverse momenta than the massless
approach.

The integrated cross-section $\sigma_{\rm meas}$ of the anti-tagged process
\epem $\to$ \epem $\DST X$ in the directly observed kinematical region $2~{\rm
GeV}<\ptdst<12~{\rm GeV}$ and $|\etadst|<1.5$ is determined to be \sigmeas. The
extrapolation of this result to the full kinematic range using Monte Carlo
simulations gives $\sigcc = \sigtot$ for the total cross-section of the
anti-tagged \epem $\to$ \epem $\rm{c\bar{c}}$  process at
$\langle\sqee\rangle=193~\mathrm{GeV}$. The first error is  statistical, the
second  is systematic  and the third  is the extrapolation uncertainty,
determined by varying the Monte Carlo parameters used in the extrapolation. 
The direct contribution is determined to be  $\sigcc_{\rm dir} = \sigdir$ while
the single-resolved contribution is $\sigcc_{\rm res} = \sigres$.

Figure~\ref{fig-sigtot} shows the total cross-section $\sigcc$  compared to
other measurements~\cite{bib-data} and to the massive NLO calculation 
of~\cite{bib-DKZZ}. Within the large band of
uncertainties, the calculation is in good agreement with most measurements.

\section*{Bottom production}
\label{sec-evsel}

A new preliminary OPAL measurement of open bottom production in photon-photon
collisions  is presented  using data collected at \epem~centre-of-mass
energies  from $\sqee=189$ to $202$~GeV corresponding to a total integrated
luminosity of about 371~pb$^{-1}$.

The selection of open beauty events proceeds in three
steps.  First, anti-tagged photon-photon events are selected.
Within these events, muon candidates are  reconstructed as a signature for 
semileptonic beauty decays.  Finally, jets are reconstructed within
the selected events, and the transverse momentum of the muon 
candidates with respect to the axis of the closest jet is computed. 

An artificial 
neural network trained for muon identification~\cite{bib-afb} is used
to enhance the purity of the muon sample.  
Events are selected if they contain exactly one muon candidate with a neural
net output $\netmu$ larger than $0.65$. 
The transverse momentum of the muons,  $\ptrel$,  is 
measured with respect to the axis of the closest jet reconstructed 
with the KTCLUS jet finding algorithm~\cite{bib-catani}. 
At least one jet with a transverse energy $\ETJET$ with respect to the 
beam axis greater than $3$~GeV and at least three tracks is required.
After this final cut $444$~events remain in this b-enriched sample.

Four contributions to the $\ptrel$ distribution have to be
taken into account: hadronic photon-photon interactions
with open b production, open c production,
the remaining hadronic photon-photon interactions (labelled `uds') and
processes other than hadronic photon-photon 
interactions (labelled `BG').

\begin{figure}[tbp]
    \includegraphics[width=.49\textwidth]{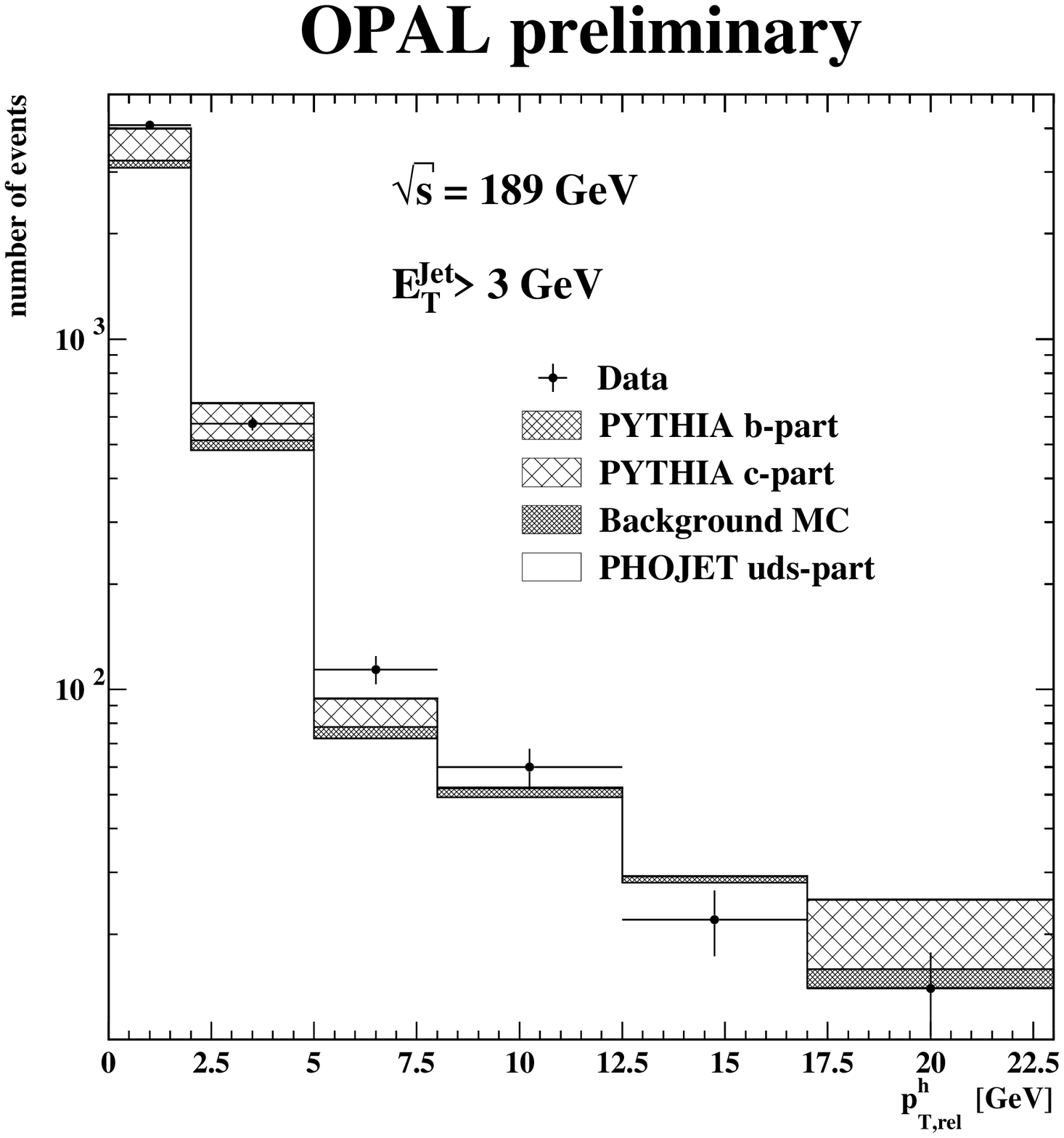}
    \hfill
    \includegraphics[width=.49\textwidth]{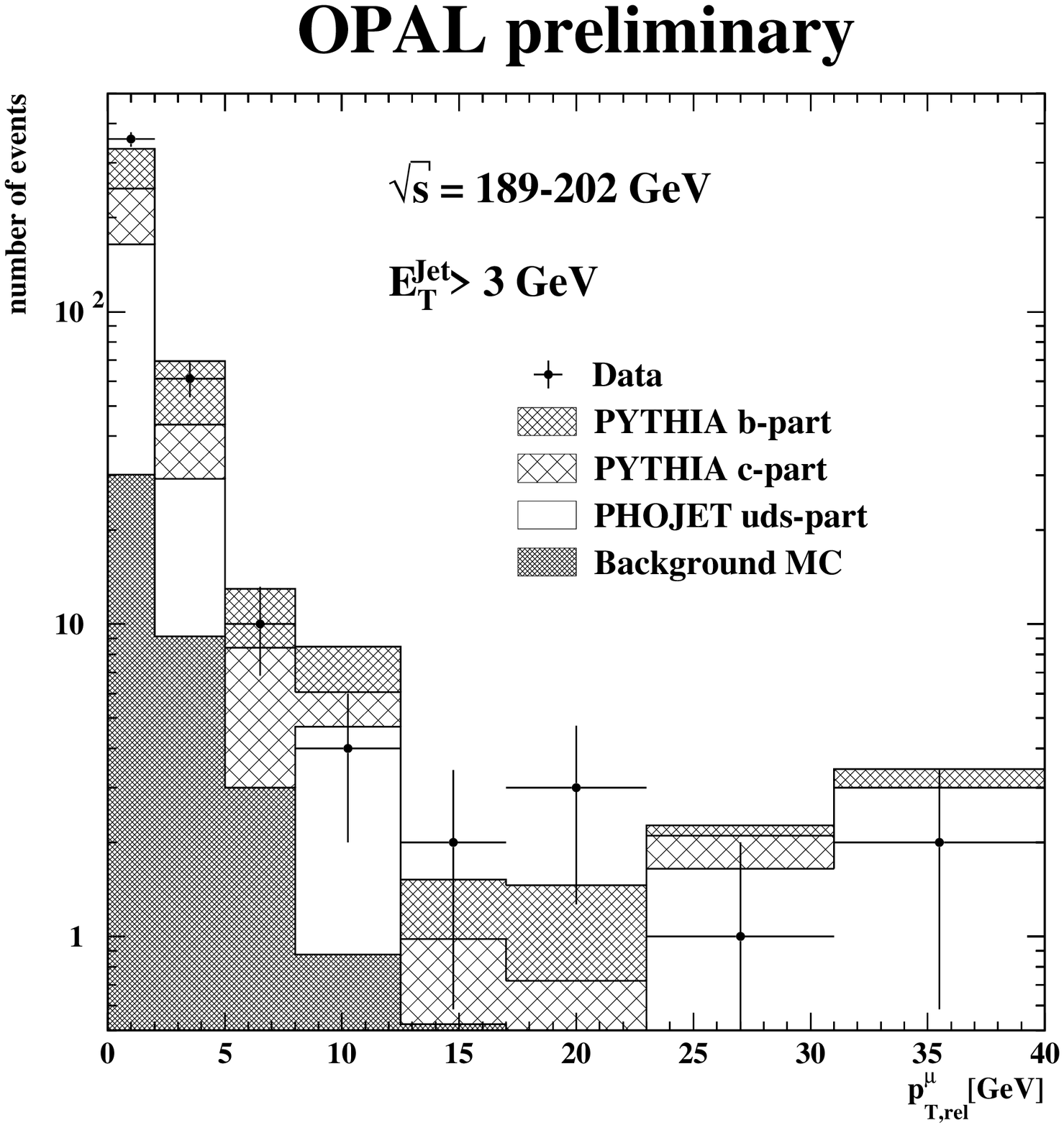}
\caption{Distribution of the transverse momentum 
with respect to the associated jet for  pions and 
kaons in the b-depleted sample (left) and for muons  
in the b-enriched sample (right).}
\label{fig-ptpion}
\label{fig-ptmuon}
\end{figure}

The number of b events is determined by a fit to the $\ptrel$ distribution
after all the other  contributions have been fixed. The previous OPAL
measurement of open charm production~\cite{bib-opaldstpaper} is used to fix the
absolute  contributions of open c production via the  direct and
single-resolved processes, where the shape is given by the PYTHIA Monte Carlo
program.  Most of the muons found in the    uds background  are due to decays
of pions and kaons into muons, therefore the rate of pion and kaon production
is measured separately.  A b-depleted sample of photon-photon events is
selected  with $\netmu<0.65$ and with pion and kaon  identification using
${\mathrm d}E/{\mathrm d}x$ measurements. The transverse momentum relative to
the jet axis for the charged hadrons in this b-depleted sample, $\pth$,  shown
in Figure~\ref{fig-ptpion}~(left), was used  to find the scaling factor of 
$1.35\pm0.03$  for the PHOJET Monte Carlo cross-section.

The predictions for processes  other than anti-tagged hadronic photon-photon 
interactions are taken entirely from Monte Carlo simulations.  The fraction
$f_{\BG}$ of background events is about $10\%$.  The main contributions to this
background are \epem annihilation events into hadrons $(54 \%)$, deep inelastic
electron-photon scattering events $(29 \%)$, production of four fermions from
other reactions than photon-photon scattering processes $(9 \%)$ and $\tau$
pairs produced in photon-photon interactions $(7 \%)$.

The final result of the one-parameter fit to the b-enriched sample,
shown in Figure~\ref{fig-ptmuon}~(right),  yields a beauty fraction 
$f_{\b}=\fb$ with $\chi^2=5.8$ (ndf=7), corresponding to $121\pm 21$ events.
The fraction of the other processes are approximately 
$f_{\c}=24\%$, $f_{\uds}=37\%$ and $f_{\BG}=10\%$.

The cross-section $\sigbb$ is determined from the measured number of events
using the branching ratios  of b hadrons into muons  taken from \cite{bib-pdg}.
The total  open b cross-section is determined to be  $\sb$. The precision of the
measurement is currently limited by the systematic errors associated with the
background from charm production, but the improved charm measurement presented
above will help to reduce this uncertainty.

\begin{figure}[tbp]
   \begin{center}
    \includegraphics[width=.5\textwidth]{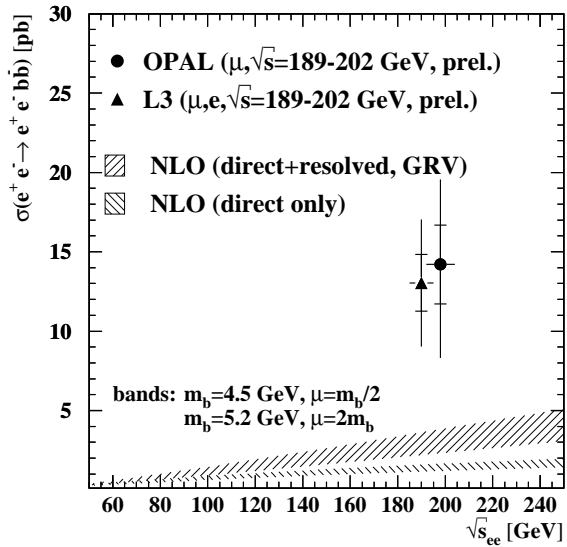}
   \end{center}
\caption{The 
total beauty production cross-section $\sigbb$ measured in the range 
$\sqee=189-202$~GeV compared to the preliminary L3 result
and the prediction of an NLO calculation.
}
\label{fig-stot}
\end{figure}

 Figure~\ref{fig-stot} shows the measured open b cross-section  compared to the
preliminary L3 result performed in the same $\sqee$ range  using semileptonic
decays of b hadrons into muons and electrons~\cite{bib-l3b}, and an NLO
calculation that uses  matrix elements for massive b quarks~\cite{bib-DKZZ},
with the direct contribution shown separately.  The prediction of the NLO
calculation for the total cross-section at  $\sqee=200$~GeV is $3.88$~pb and
$2.34$~pb for a b quark mass of $4.5$ GeV and $5.2$~GeV, respectively,
significantly lower than both measurements.
A recent measurement of the open beauty cross-section in photon-proton 
collisions is also about 2.5 standard deviations higher than expected  from an
NLO calculation~\cite{bib-bhera}.

\section*{Conclusion}

The inclusive production of $\DSTPM$ mesons  has been measured in  anti-tagged
photon-photon collisions using the OPAL detector at LEP. The contribution of
the direct process  in the  kinematical region $\ptdst>2$~GeV and
$|\etadst|<1.5$ is determined from the  $\xt$ distribution to be  $r_{\rm dir}
= (44\pm 6)\%$.

The measured differential cross-section as a function of 
the $\DST$ transverse momentum
 and pseudorapidity is  compared to 
 NLO calculations using the massless and  the massive
 approaches, and despite the low values of $\ptdst$ studied 
 the massless calculation is in good agreement with the data, while
the massive calculation  underestimates them for lower values of
 $\ptdst$.

The cross-section of the anti-tagged \epem $\to$ \epem $\DST X$ process  is
measured in the restricted kinematical range of  $\ptdst>2$~GeV and
$|\etadst|<1.5$ to be \sigmeas. The extrapolation of this result to the total
charm cross-section  introduces large uncertainties both on the theoretical and
experimental side, therefore at present a comparison in the restricted
kinematic range is preferred.


The open b cross-section in photon-photon events has been measured using the
semi-leptonic decays of b hadrons into muons. The spectrum of the transverse
momentum of the muons with respect to the closest jet axis  is fitted  after
all background contributions have been fixed independently, leading to a
measured total  cross-section   of  $\sb$, in agreement with the preliminary 
L3 measurement. However, the NLO QCD calculation underestimates the measured
total cross-section by about 2 standard deviations.

\end{document}